\documentclass[12pt,preprint]{aastex}
\begin{document}

\title{\emph{HST} SPECTROPHOTOMETRY AND MODELS FOR SOLAR ANALOGS}

\author{R.~C.\ Bohlin\altaffilmark{1}
\altaffiltext{1}{Space Telescope Science Institute, 3700 San Martin Drive,
Baltimore,  MD 21218; bohlin@stsci.edu}}
\slugcomment{To appear in AJ 2010 April}
\received{2009 Dec 29} 
\accepted{2010 Feb 10}
%%% \accepted{xxx}

\begin{abstract}

Absolute flux distributions for seven solar analog stars are measured from 0.3
to 2.5~$\mu$m by \emph{HST} spectrophotometry. In order to predict the longer wavelength mid-IR fluxes that are required for \emph{JWST} calibration, the \emph{HST} SEDs are fit with Castelli \& Kurucz model atmospheres; and the results are compared with fits from the MARCS model grid. The rms residuals in 10 broad band bins are all $<$0.5\% for the best fits from both model grids. However, the fits differ systematically: The MARCS fits are 40--100~K hotter in $T_\mathrm{eff}$, 0.25--0.80 higher in $\log g$, 0.01--0.10 higher in $\log z$, and 0.008--0.021 higher in the reddening $E(B-V)$, probably because their specifications include different metal abundances.  Despite these differences in the parameters of the fits, the predicted mid-IR fluxes differ by only $\sim$1\%; and the modeled flux distributions of these G~stars have an estimated ensemble accuracy of 2\% out to 30~$\mu$m.

\end{abstract}

\keywords{stars: atmospheres --- stars: fundamental parameters --- stars:
 individual (HD209458, P041C, P177D, P330E, C26202, SF1615+001A, SNAP-2) --- techniques: spectroscopic}

\section{Introduction}

The \emph{James Webb Space Telescope} (\emph{JWST}) requires flux standards in the 0.8--30~$\mu$m region. To define reference absolute flux distributions at the longer wavelengths, stellar model SEDs can be anchored to precision \emph{Hubble Space Telescope} (\emph{HST}) spectrophotometry at 0.3--2.5~$\mu$m and extrapolated into the mid-IR. A variety of stellar types helps guard against systematic effects in the modeling and extrapolation process, while several standards of each type provide a statistical reduction of the random errors in the measured fluxes and in the fitting process. Three types: white dwarfs (WDs), A~stars, and 
G~stars are chosen because of the precision of the existing \emph{HST} spectrophotometry from the STIS ($R\sim1000$) and NICMOS ($R\sim200$) instruments. For this paper, the SEDs of seven G~stars are determined by STIS spectra in the wavelength region below 1~$\mu$m, while the longer wavelengths to 2.5~$\mu$m are based on NICMOS grism data. These seven
SEDs are archived in the CALSPEC database of \emph{HST} flux
standards\footnote{http://www.stsci.edu/hst/observatory/cdbs/calspec.html/} as \emph{*\_stisnic\_003.fits} and include vectors of estimates of the statistical and systematic uncertainties, the resolution (FWHM), the data quality, and the total exposure time, in addition to the wavelength and flux vectors.

The pure hydrogen WD stars G191B2B, GD71, and GD153 are the primary calibrators for the \emph{HST} spectrophotometry; and these three reference SEDs show internal consistency to better than $\sim$1\% (Bohlin 2007, Bohlin, Riess, \& de~Jong 2006; Bohlin, Dickinson, \& Calzetti 2001.) The fainter WD LDS749B fits the continuum of its pure helium model to a similar excellent accuracy (Bohlin \& Koester 2008). Nine A~star standards have been published by Bohlin \& Cohen (2008) and are also observed by the Infrared Array Camera (IRAC) on the \emph{Spitzer Space Telescope} (Reach et~al.\ 2005). Seven of these $R\sim200$ resolution NICMOS flux distributions fit their A~star model SEDs to 1\%. 

In addition to the Castelli \& Kurucz (2004 CK04) grid used by Bohlin \& Cohen
(2008) to model the A~stars, the MARCS\footnote{http://www.marcs.astro.uu.se} grid (Gustafsson et~al.\ 2008) is available for models cooler than $T_\mathrm{eff}=8000$~K. The subset of MARCS grids with plane-parallel structure and solar metal abundances is used to model the G~stars. For the G~star spectrophotometry discussed in this paper, the broad band \emph{HST} fluxes and both sets of models all agree to 0.5\%. The ability to provide a consistent set of model flux distributions that fit the measured SEDs of WDs, A~stars, and G~stars lends confidence to this system of high precision \emph{HST} standard stars.

\section{Solar Models}

Kurucz (2004)\footnote{http://kurucz.harvard.edu/stars/sun} has produced a
finely sampled solar model SED (fsunallp). Figure~1 compares both this special solar spectrum and the MARCS model to the same CK04 model interpolated in the CK04 grid. All three model SEDs are for the same effective temperature $T_\mathrm{eff}=5777$~K, surface gravity $\log g=4.44$, metallicity [M/H]\,$=\log z=0$, and the turbulent velocity $v_{\rm turb}=2$~km/s, except that the Kurucz fsunallp model has $v_{\rm turb}=1.5$~km/s. Shortward of 0.45~$\mu$m, where the line blanketing is severe, the MARCS model differs from the CK04 model by more than 5\% in the bottom panel of Figure~1, and the Kurucz (2004) and CK04 models
differ by up to 3\% in the top panel. In this region of severe line blanketing,
the models are most sensitive to the input parameters. For example, changing
only the $v_{\rm turb}$ from two to one km/s increases the mean flux in the
0.30--0.37~$\mu$m range by $\sim$5\%, while affecting the fluxes longward of 0.40~$\mu$m by $<$1\%.

Longward of 0.6~$\mu$m, the Kurucz (2004) and CK04 models differ smoothly by up to 2\% in the top panel; however, the ratios in the two panels of Figure~1 match to $\sim$1\%. This match between the Kurucz (2004) and MARCS models suggests that the CK04 models may have a slight IR excess with respect to the $V$~band. Furthermore, direct comparisons of the CK04 models with the \emph{HST} observations also show similar residuals in the 1--2.5~$\mu$m region. Thus, for solar analog stars with $T_\mathrm{eff}$ within a few hundred degrees of the Sun, the first-order correction in the top panel should be an improvement to the CK04 grid models longward of 1~$\mu$m. In the 4.2--7~$\mu$m region where the
Kurucz (2004) model overestimates the strength of the main CO band absorption (Kurucz, private communication), no correction is made to the CK04 grid models. Whenever CK04 is mentioned below, this correction procedure that is based on the Kurucz special solar model is implicit from 1--4.2~$\mu$m and longward of 7~$\mu$m. The correction is smooth and does not exceed 0.5\% beyond 7~$\mu$m.

\section{Model Parameters for the Best Fit to \emph{HST} Fluxes}

Following the technique of Bohlin \& Cohen (2008), values for $T_\mathrm{eff}$, $\log g$, $\log z$, and the selective dust extinction between the $B$ and $V$~bands, $E(B-V)$, are derived by fitting the \emph{HST} observational SEDs with model atmospheres from the grid of Castelli \& Kurucz (2004 CK04). Figure~2 compares the observations to the best fitting CK04 models. Similarly, Figure~3 shows the ratio of the observations to the best fit from the MARCS grid (Gustafsson, et~al.\ 2008). Each reddened model is normalized to the observed flux averaged over 0.6--0.9~$\mu$m. The reddening curve used to correct for the extinction as a function of wavelength and $E(B-V)$ is from Cardelli, Clayton, \& Mathis (1989, CCM) at wavelengths shorter than 2~$\mu$m and is from Chiar \& Tielens (2006) at
longer wavelengths. Both the  level and slope of the two extinction curves are
matched at 2~$\mu$m. 

In Figures~2--3, the ratios of the observations to the models appear as small
circles at a resolution of $R=100$. The large scatter in these narrow band small circles at the shorter wavelengths demonstrates the Kurucz (2005) claim that exact model fits are currently hopeless because of poor and incomplete atomic line data and because of elemental abundance uncertainties, especially in regions of strong line blanketing. Typically for solar type stars, the model computations show a line blanketing that exceeds 20\% below 0.45~$\mu$m. However, averages over broad bands should be accurate because of the statistical nature of the uncertainties of the absorption line strengths. For the fainter stars in the three top panels, the signal to noise of the observations is poor; and much of the $R=100$ fine structure is just noise. For the four brighter stars, Figure~4 illustrates the source of the fine structure in one narrow wavelength region. Longward of $\sim$0.42~$\mu$m, the CK04 models in Figure~4 are low, which causes the peaks around 0.43~$\mu$m that often rise above 1.05 in Figure~2. Similarly, the MARCS models are systematically higher than the observations near 0.41~$\mu$m, which causes the narrow dips at that wavelength in Figure~3. The use of models for narrow band calibrations is problematic because of deviations from the true observed flux and because of the coarse wavelength grid for the CK04 models. For calibrations in narrow bands with a spectral resolution comparable to the observations, the finite resolution of the observations may also cause errors to the extent that spectral features
fall in the band pass and to the extent there is noise in the observed flux.
Whenever possible, narrow band filters and spectroscopic sensitivities are best
measured using a standard star with minimal spectral structure in the region of interest. The effects of the resolution and noise can be alleviated by fitting a smooth function to the sensitivity as a function of wavelength in the case of a spectrograph calibration; but narrow band filter calibrations suffer errors directly from all the above problems.

Table~1 lists the broad band regions used to minimize the rms scatter in the
fitting procedure. The averages over the first two broad bands show opposite
behavior as a function of gravity, in the sense that the flux decreases at
0.300--0.372~$\mu$m and increases at 0.372--0.455~$\mu$m as $\log g$ increases. These two bands provide a strong constraint on the derived values for $\log g$ in the temperature regime of the program solar analog stars. For example, an increase of $\sim$10\% in the first to the second broad band flux ratio represents a change of $+1$ in $\log g$. Because the uncertainties in the NICMOS data are larger than for STIS, the NICMOS bands are wider. Furthermore, an extra uncertainty of 1--2\% exists where the NICMOS non-linearity correction of Bohlin, Riess, \& de Jong (2006) is the largest at 1--1.3~$\mu$m. Therefore, this 1--1.3~$\mu$m region is not used to constrain  the model fits. The large open circles in Figures~2--3 represent the average ratio in each of these 10 continuum regions, while the rms value appearing for each star represents the scatter about the fit in the 10 broad bands. There is only one NICMOS observation of P177D longward of 1.9~$\mu$m, so the last broadband is not used for the fitting of P177D. The granularities of the fits are 20~K, 0.05, 0.01, and 0.001 in $T_\mathrm{eff}$, $\log g$, $\log z$, and $E(B-V)$, respectively.

Table~2 summarizes the results of the best fits for the two sets of models. The residual broad band rms values are all $<$0.5\% for both the CK04 and MARCS model fitting.

\section{Comparison of CK04 and MARCS Results}

There are systematic differences in the models: the MARCS and the CK04 best fits differ by up to +100~K in $T_\mathrm{eff}$, +0.8 in $\log g$, +0.10 in [M/H], and +0.021 in $E(B-V)$, even though both grids are computed for plane parallel structure and with $v_{\rm turb}=2$~km/s. These differences are probably due largely to different adopted abundances. The Marcs models use the relative abundances and the $Z=0.012$ fractional abundance by mass of chemical elements heavier than helium from Grevesse, Asplund, \& Sauval (2007), while CK04 adopt the abundances of Grevesse \& Sauval (1998) with $Z=0.017$. 

The low values from the CK04 grid for $\log g$ with respect to the MARCS results and with respect to the 4.44 solar value may indicate that the metallicities adopted for the MARCS models are more realistic in the heavily line blanketed 0.3--0.455~$\mu$m region that is of primary importance for the determination of $\log g$. Whatever the true cause of the low values for $\log g$ for the CK04 fits, there is little effect on the mid-IR flux, which is the main goal of this work. For example in the worst case of P177D with the lowest $\log g$ of 3.60, increasing $\log g$ to the canonical solar value of 4.44 makes large changes of a few percent in the two shortest wavelength broad bands, as expected. The 0.3--0.372~$\mu$m ratio drops by $\sim$6\%, while the next band at 0.372--0.455~$\mu$m increases by $\sim$3\%. In these heavily line-blanketed regions, such deviations can be explained by abundance errors and uncertainties in the atomic physics used by CK04. However, at the longer wavelengths, where the goal is to accurately predict the flux, the maximum difference between the $\log g$ of 3.60 and 4.44 models is 0.2\%. The values for $T_\mathrm{eff}$ and the reddening $E(B-V)$ are much more important than $\log g$ for determining the IR fluxes with respect to the 0.6--0.9~$\mu$m normalization region.

The MARCS models are indexed by a quantity $\log z=$\,[Fe/H], which represents the logarithm of the number of iron atoms relative to hydrogen by number; and [Fe/H]\,$=0.0$ indicates a solar abundance. The upper~($Z$) and lower~($z$) cases must not be confused; but as long as models with enhanced alpha elements are not considered, then the fraction of metals by mass, $Z$, scales directly with the total metal number density relative to the Sun, $\log z=$\,[Fe/H]\,=\,[M/H], where M represents the total metallicity. A quirk of the MARCS grid is that for [Fe/H]\,$<0$, a solar relative metallicity among the heavy elements is classified as ``alpha-poor,'' while models with enhanced oxygen and alpha elements with respect to iron are denoted as ``standard.'' Hence, for the six stars with metallicities lower than solar, the ``alpha-poor'' MARCS grid is utilized.

Despite the differences in the models, the comparison in Figure~5 of the best
fitting CK04 and MARCS models for each star typically agree to $\sim$1\% in
broad continuum bands. The systematically higher temperatures for the MARCS fits are partially compensated by higher reddenings. Unfortunately, this comparison between CK04 and MARCS extends only to 20~$\mu$m, which is the long wavelength limit of the MARCS grid. Because \emph{JWST} requires flux standards to 30~$\mu$m, the CK04 models are the baseline, while the agreement with the independent MARCS grid sets a lower limit of $\sim$1\% to the uncertainty in the modeled IR SEDs.

In addition to this systematic uncertainty in the G~star models, the systematic
uncertainty in the SEDs of the primary pure hydrogen WDs that are the basis for the \emph{HST} flux scale must be included in the error analysis. The error model of Bohlin (2003) assigns a 1\% uncertainty in the flux relative to the 0.55~$\mu$m flux for these WDs longward of 1.5~$\mu$m. For models in the range of the results of Table~2, a delta of 18~K, e.g., for models with
$T_\mathrm{eff}=5800$ vs.\ 5818~K, the change in the 1.5/0.55~$\mu$m flux ratio is 1\%. However, this possible error in the \emph{HST} flux scale causes little extra error in the extrapolated G~star fluxes, because the ratio of these two $\sim$5800~K models is rather flat beyond 1.5~$\mu$m, increasing to an error of only 1.2\% at 30~$\mu$m. Thus, combining the $\sim$1\% uncertainty in the G~star models with a $\sim$1\% uncertainty in the primary WD models, a limit of 2\% to the systematic uncertainty for the modeled mid-IR fluxes relative to the $V$~band is conservative, except in the unlikely case that both model grids have errors in the mid-IR that are large \emph{and} of similar amount.

The CK04 fits often show somewhat lower \emph{rms} residuals in Table~2; and the CK04 fits are  usually closer to the canonical solar values of $T_\mathrm{eff}=5777$~K, while the MARCS models are closer to the solar $\log
g=4.44$. A difference common to all seven stars in Figure~5 is the rise of
$\sim$2\% from 4~$\mu$m to the maximum of the CO fundamental band strength at 4.6~$\mu$m, where the modeling is severely challenged by the difficulties of the molecular physics and the tenuousness of the solar atmosphere at the height of the temperature minimum where CO is most abundant. Precision \emph{JWST}/NIRSPEC spectrophotometry of the 
G~stars relative to the WD and A~stars should determine the true strength of the CO bands, which will enable better estimates of the SEDs of the 
G~type standards in this region of greatest uncertainty.

\section{HD209458}

HD209458 has a planetary transit that reduces the flux by $\sim$1.6\%
(Wittenmyer et~al.\ 2005). However, the length of the transit is only $\sim$3
hours from phase $\sim$0.98--1.02 in the 3.5 day orbital period.  The ephemeris is precisely known; and any observation of HD209458 can be checked for dimming caused by a transit. The STIS and NICMOS observations utilized here are all obtained outside of transit. The precise period is $P=3.52474554$ days with an uncertainty of 0.016~s, so that an uncertainty of the time of zero phase is less than one minute for 36 years after 2003, when the time of zero phase was measured precisely. The heliocentric phase at any time $T$ is the fractional part of $(T-T_\mathrm{o})/P$, where $T_\mathrm{o}=52,854.82545$ is the Reduced Julian
Date, i.e., with 2,400,000 days subtracted. For example, the STIS observation at 2001 October 28 22:13:24 UT is at $T=52211.4260$~JD or at a phase of $-0.54$, i.e., $+0.46$. For precise estimates, the difference in light travel time to the Earth instead of the Sun must be accounted. For the ecliptic latitude of $28.7^{\circ}$ for HD2090458, this time difference is in the range $\pm$7.3 minutes, i.e., $\pm$0.001 in phase.

During secondary eclipse at phase 0.5, Knutson, et~al.\ (2008) report flux
decreases of 0.09\%, 0.21\%, 0.30\%, 0.24\%, and 0.26\% at 3.6, 4.5, 5.8, 8.0, and 24~$\mu$m, respectively, while Rowe et~al.\ (2006) constrain the planetary flux contribution to $<$0.01\% in the 0.4--0.7~$\mu$m band. Thus, the model predictions for the SED of HD209458 are in error by small fractions of a percent as a function of wavelength due to the contribution from the hot Jupiter planet. Outside of eclipse, Cowan et~al.\ (2007) constrain the variation with phase to be $<$0.15\% at 8~$\mu$m.

Also, the temperatures for HD209458 of 6080 and 6160~K in Table~2 are near those summarized by Wittenmyer et~al.\ (2005), especially the $T_\mathrm{eff}$ value of 6099 $\pm$44~K of Fischer \& Valenti (2005), who find $\log g=4.38$, and [Fe/H]\/$=0.01$. The special model with $T_\mathrm{eff}=6100$~K and $\log g=4.38$ for HD209458\footnote{http://kurucz.harvard.edu/stars/hd209458/} constructed by Kurucz in 2005 agrees with the best fit CK04 model to within about 2\% longward of 0.4~$\mu$m, when the two models are normalized at 0.6--0.9~$\mu$m. 

\section{The Sun as a Solar Analog Star}

Figure~6 shows the comparison between the solar observations and the models, in analogy with the stellar observations in Figures~2--3. The flux of the Sun is from Thuillier et~al.\ (2003, Th03), where observations from the ATLAS and EURECA missions with the Space Shuttle are referenced to the Heidelberg Observatory blackbody absolute-flux standard. Rieke et~al.\  (2008) have adopted the Th03 flux distribution for their solar SED below 2.4~$\mu$m. The agreement between the shapes of the observed and modeled SED does not achieve the 1\% goal, as illustrated in the bottom two panels, where the Th03 measurements differ from both models by $\sim$4\% near 2~$\mu$m. Thus, either the measured Th03 fluxes or both special solar models have the wrong $K$-band to $V$-band flux ratio. The Th03 uncertainty at 2~$\mu$m is quoted as 1.3\% in Table~III of Th03
but grows to ``$\sim$2\%'' in the Conclusions section, which also quotes a 2--3\% uncertainty below 0.85~$\mu$m. Thus, an error of 4\% in the ratio of the 2 to 0.55~$\mu$m Th03 fluxes is conceivable. If the Th03 solar SED does actually have errors as large as 4\%, then the \emph{HST} SEDs may be more accurate with their estimated uncertainty of 2\% in relative flux.

Conversely, if the models have small systematic errors, for example, in the abundances, then somewhat different models will fit the TH03 fluxes better. In the upper two panels, the best fits from the respective grids suggest that a model with a somewhat lower metallicity may be required for the Sun, i.e., $\log z= -0.29$ and $-0.23$ for the best fits from the CK04 and MARCS grids,
respectively.

At the longer wavelengths, Figure~7 illustrates the ratio of three other
candidate solar SEDs to the baseline CK04 model. In the bottom panel, the
agreement of the MARCS SED with CK04 is generally within 1\%, except in the problematic CO band around 4.6~$\mu$m. In the middle panel from 0.5--2.4~$\mu$m, the comparison with Rieke et~al.\ (2008) is the same as shown for the comparison with Th03 in the bottom panel of Figure~6, except for the more drastic smoothing. Longward of 2.4~$\mu$m, the agreement of CK04 with the model used by Rieke et~al.\ (2008) is excellent, i.e., within the estimated uncertainty of 2\%. In the top panel, the comparison is with the best fit CK04 model from Figure~6. The ratio of this cooler model SED with lower gravity and lower metallicity to the baseline CK04 model rises to 1.06 at 30~$\mu$m. Such a large discrepancy as 1.06 seems unlikely given the excellent agreement among the standard CK04, MARCS, and Rieke models longward of 2.4~$\mu$m. Thus, the most likely conclusion seems to be that the ratio of the Th03 flux at 2~$\mu$m to the Th03 flux shortward of 1~$\mu$m is too high by $\sim$4\%. The only other possibility is that all the models are wrong by a similar amount in their $V$ to $K$~band and 
$K$~band to 30~$\mu$m flux ratios.

\section{Summary}

Absolute flux standards with spectral types of G are required for \emph{JWST} calibration. None of the seven \emph{HST} stellar G-type stars is a true ``solar analog'' in the sense that the shape of the SED matches either the Th03 solar flux measurements or the standard Kurucz (fsunallp) or MARCS solar model to 1\% over the whole 0.3--2.5~$\mu$m observed range. However, STIS and NICMOS spectrophotometry measure the flux distributions of these seven G~stars, which fit CK04 and MARCS model atmospheres from 0.3--2.4~$\mu$m within an rms scatter of $<$0.5\% in broad wavelength bands. These models are normalized to the \emph{HST} SEDs and are used to extend the measured fluxes to longer wavelengths. The seven composite observed plus model SEDs are archived in the CALSPEC database of \emph{HST} flux standards\footnote{http://www.stsci.edu/hst/observatory/cdbs/calspec.html/} as
\emph{*\_stisnic\_003.fits}. 

Emission from dust rings like the one around Vega often contribute added mid-IR flux; but \emph{JWST} observations can identify discrepantly large cases of excess emission longward of 10~$\mu$m, if no other mid-IR data are available before \emph{JWST} operations begin. A separate paper with additional authors is planned, which will compare the SEDs of some of these G~stars and a set of \emph{HST} WDs and A~stars with the revised \emph{Spitzer} calibration of Rieke et~al.\ (2008).

\acknowledgments

R.~Gilliland, K.~Gordon, J.~Kalirai, R.~Kurucz, and the Referee supplied
extremely helpful comments on drafts of this paper. D. Lindler provided the IDL routine \emph{model\_int.pro} and the databases for interpolating in the model grids. Support for this work was provided by NASA through the Space Telescope Science Institute, which is operated by AURA, Inc., under NASA contract NAS5-26555. This research made use of the SIMBAD database, operated at CDS, Strasbourg, France.

\begin{deluxetable}{c}
\tablewidth{0pt}
\tablecaption{Broad Bands for Fitting Models}
\tablehead{
\colhead{Wavelength Range ($\mu$m)}
}
\startdata
0.300--0.372\\
0.372--0.455\\
0.455--0.550\\
0.550--0.650\\
0.650--0.750\\
0.750--0.850\\
0.850--1.000\\
1.300--1.550\\
1.550--1.900\\
1.900--2.400\\
\enddata
\end{deluxetable}

\begin{deluxetable}{lllrcccrcccccrcc}
\rotate
\tabletypesize{\scriptsize}
\tablewidth{0pt}
\tablecolumns{17}
\tablecaption{Solar Analog Stars}%1
\tablehead{
\colhead{Star} &\colhead{R.A.} &\colhead{Decl.} &\colhead{$V$}
&&\colhead{$T_\mathrm{eff}$} &\colhead{$\log g$} &\colhead{[M/H]} &\colhead{$E(B\!-\!V)$} &\colhead{rms (\%)}
&&\colhead{$T_\mathrm{eff}$} &\colhead{$\log g$} &\colhead{[M/H]} &\colhead{$E(B\!-\!V)$} &\colhead{rms (\%)}\\
&\multicolumn{1}{c}{J2000} &\multicolumn{0}{c}{J2000} &&\multicolumn{5}{c}{CK04} &&\multicolumn{5}{c}{MARCS}
}
\startdata
P041C       &14 51 58.19 &+71 43 17.3 &12.01    &&5960 &3.95 &0.02 &0.027 &0.31    &&6020 &4.65 &0.03 &0.038 &0.34\\
P177D       &15 59 13.59 &+47 36 41.8 &13.48    &&5780 &3.60 &$-0.17$   &0.031 &0.24    &&5860 &4.30 &$-0.08$  &0.049 &0.29\\
P330E       &16 31 33.85 &+30 08 47.1 &13.01    &&5820 &4.00 &$-0.30$   &0.035 &0.28    &&5920 &4.60 &$-0.20$  &0.056 &0.41\\
HD209458    &22 03 10.8  &+18 53 04   &7.65  &&6080 &4.00 &$-0.10$   &0.005 &0.15    &&6160 &4.45 &$-0.04$  &0.021 &0.24\\
C26202      & 3 32 32.88 &$-$27 51 48.0 &16.64    &&6100 &4.30 &$-0.55$   &0.035 &0.34    &&6200 &4.55 &$-0.48$  &0.054 &0.47\\
SF1615+001A &16 18 14.23 &+00 00 08.4 &16.75    &&5800 &4.10 &$-0.78$   &0.104 &0.29    &&5840 &4.45 &$-0.73$  &0.112 &0.43\\
SNAP-2      &16 19 46.13 &+55 34 17.7 &16.2\phn     &&5740 &4.05 &$-0.36$   &0.036 &0.34    &&5800 &4.85 &$-0.31$  &0.048 &0.46\\
\enddata
%\tablenotetext{a}{Simbad. Simbad has A2 for HD~165459.}
\end{deluxetable}
%HD209458 &\phn6.994\rlap{\tablenotemark{a}} &\phn6.864\rlap{\tablenotemark{a}} &\nodata &\nodata &9397 &4.18 &A1V &0.09 &&8600 &4.20 &$-1.5$ &0.017 &A4V &0.12 &0.13 &0.01\\

\clearpage
% convert fig4.pdf fig4.eps
%\includegraphics*[scale=.75,trim=0 0 0 0,angle=90]{fig5.eps}
\begin{figure}  
\centering  
\includegraphics[width=\textwidth]{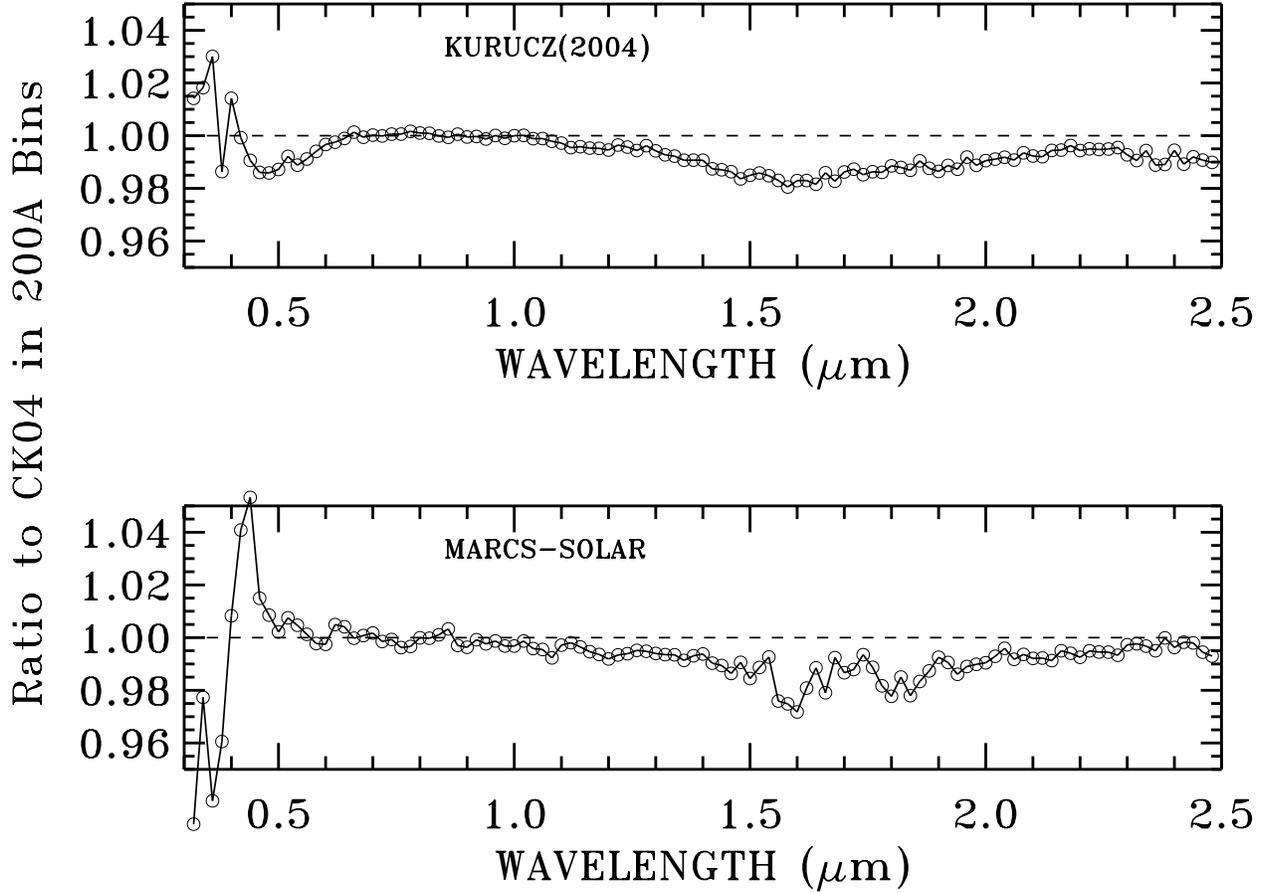}
\caption{The ratio of the \emph{fsunallp} Kurucz (2004) solar model (\emph{upper panel}) and the MARCS (\emph{lower panel}) to the $T_\mathrm{eff}=5777$~K, $\log g=4.44$ model interpolated from the CK04 grid. Longward of 1~$\mu$m, the ratio in the upper panel is used to correct all models interpolated from the CK04 grid.} 
\end{figure}

\begin{figure} 
\centering 
\includegraphics[height=7in]{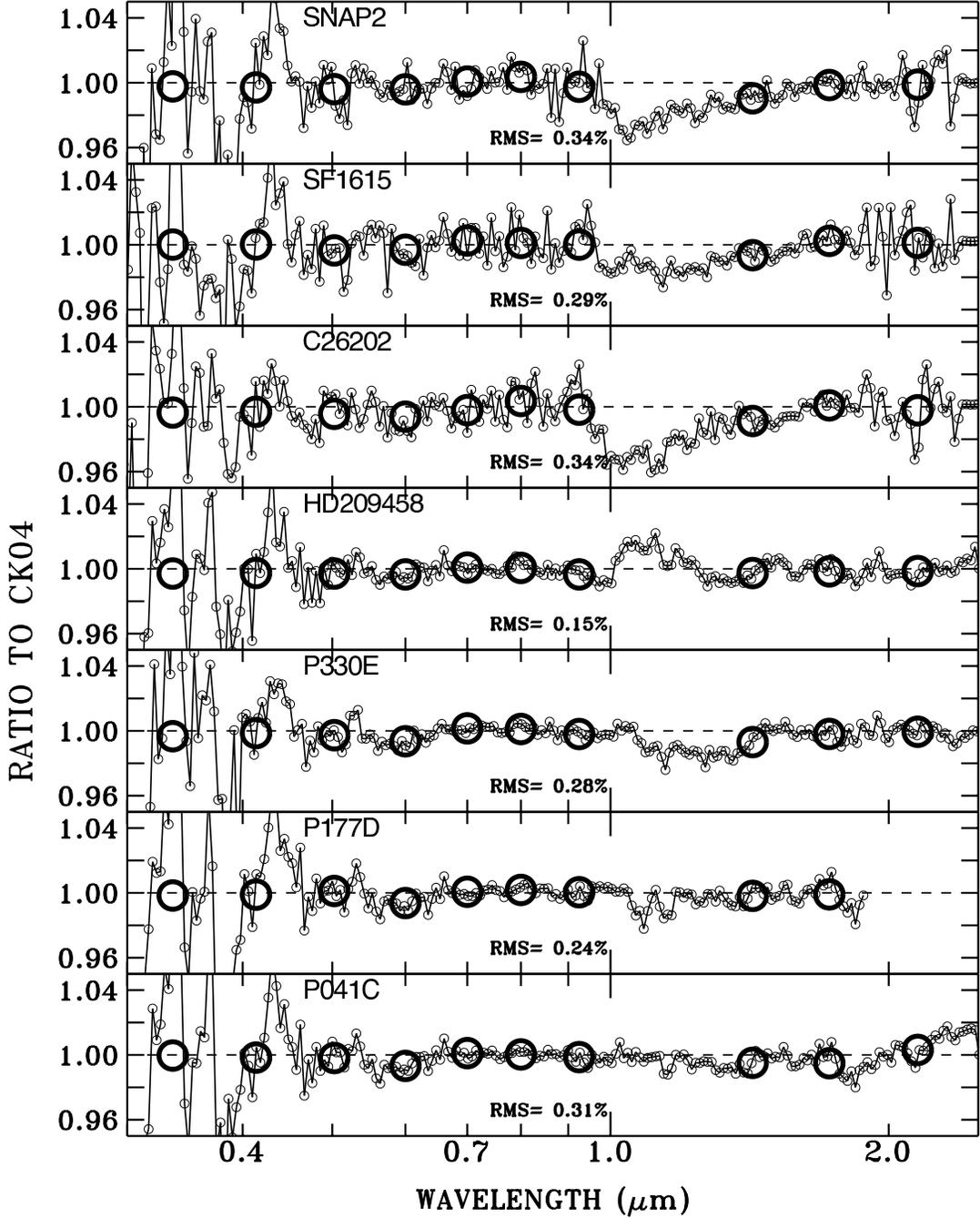}
\caption{\baselineskip=16pt
The ratio of the \emph{HST} fluxes to the best fitting CK04 model SED at $R=100$ (\emph{small circles}) and in the broad bands of Table~1 (\emph{large circles}). The denominators are interpolated from the CK04 model grid using the IDL routine \emph{model\_int.pro}. The stellar parameters for the CK04 models are in Table~2. The models are normalized to the \emph{HST} fluxes over the 0.6--0.9~$\mu$m range. Before binning to $R=100$, the models are smoothed to the resolution of the observations.} 
\end{figure}

\begin{figure}
\centering
\includegraphics[height=7in]{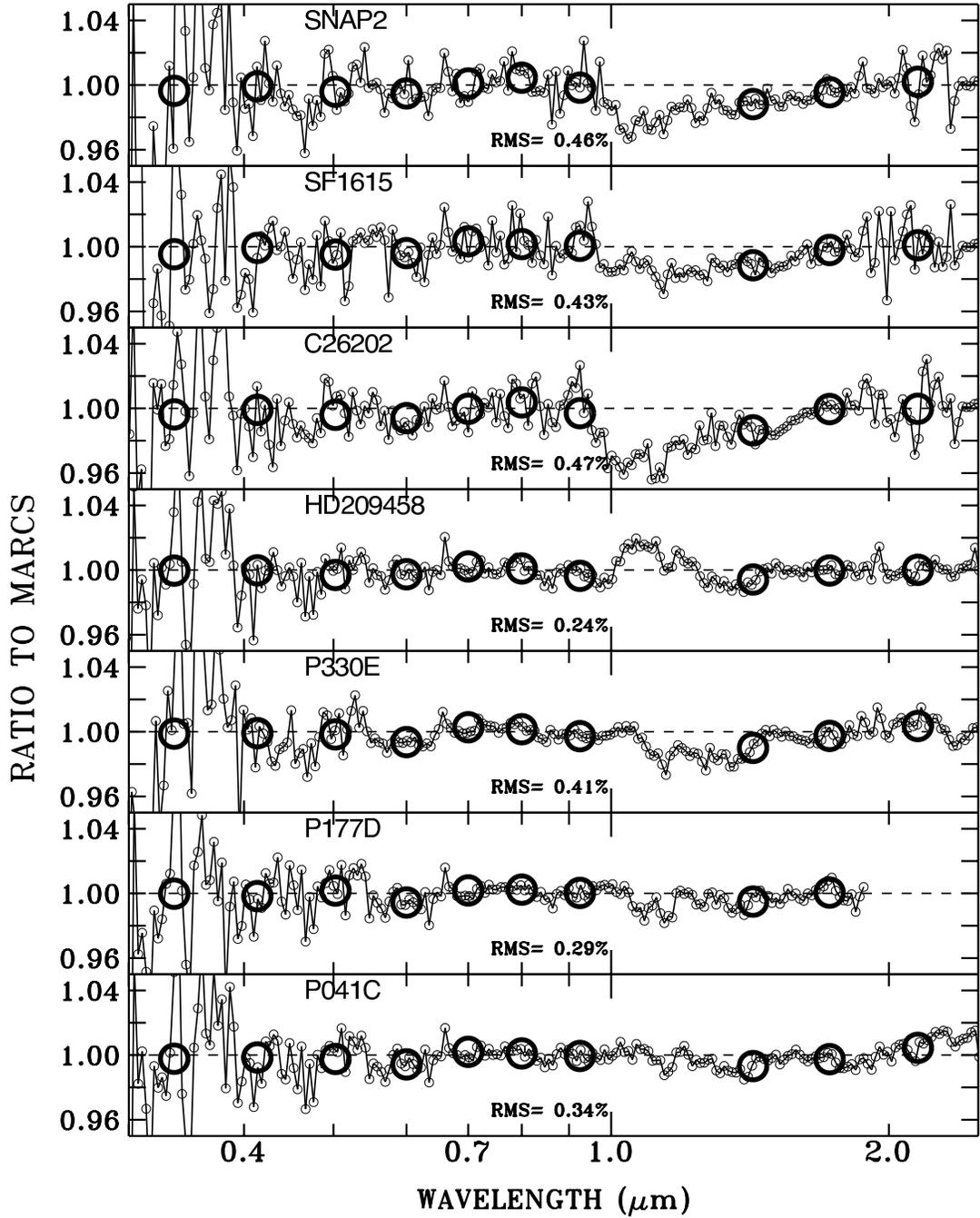}
\caption{As in Figure 2, except that the denominators are interpolated from the MARCS grid.} 
\end{figure}

\begin{figure}
\centering
\includegraphics[height=6.5in]{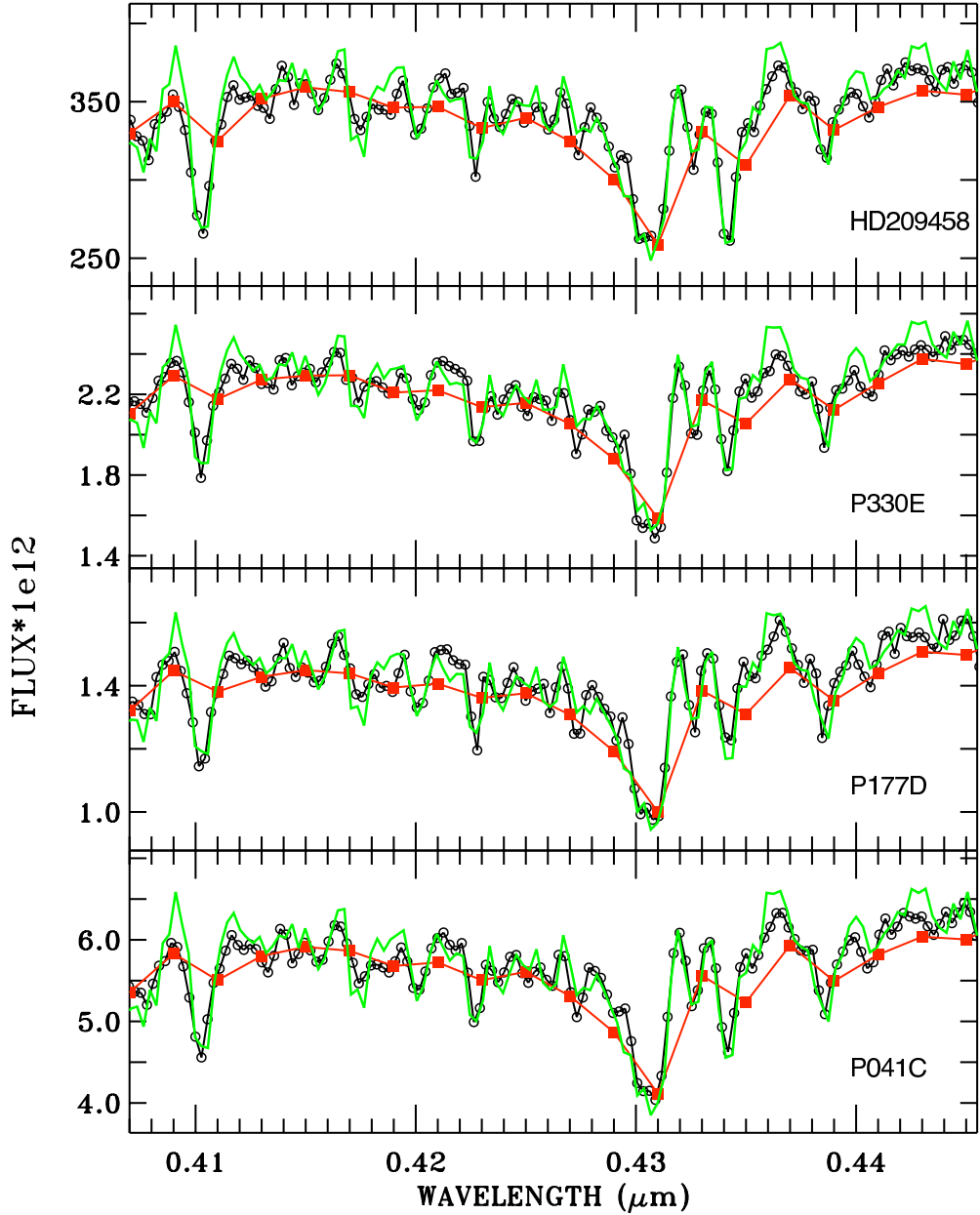}
\caption{\baselineskip=16pt
Detailed comparison of the STIS absolute fluxes (\emph{black~circles}) for the four brightest stars with their best fitting CK04 (\emph{red~squares}), and MARCS (\emph{green}) models in a narrow wavelength region of 0.41--0.44~$\mu$m. The CK04 SED is shown at the resolution of the sparse grid of 1221 points that covers the full 0.009--160~$\mu$m wavelength range, while the finely sampled MARCS models are smoothed to match the $R\sim1000$ resolution of the black circle STIS data. The bulk of the fine structure in Figures 2--3 is \emph{not} due to missing strong absorption lines in the models; but instead, the few percent differences between
observation and theory are caused by small errors in the total stength of the
heavy line-blanketing.}
\end{figure}

\begin{figure}
\centering
\includegraphics[height=\textwidth]{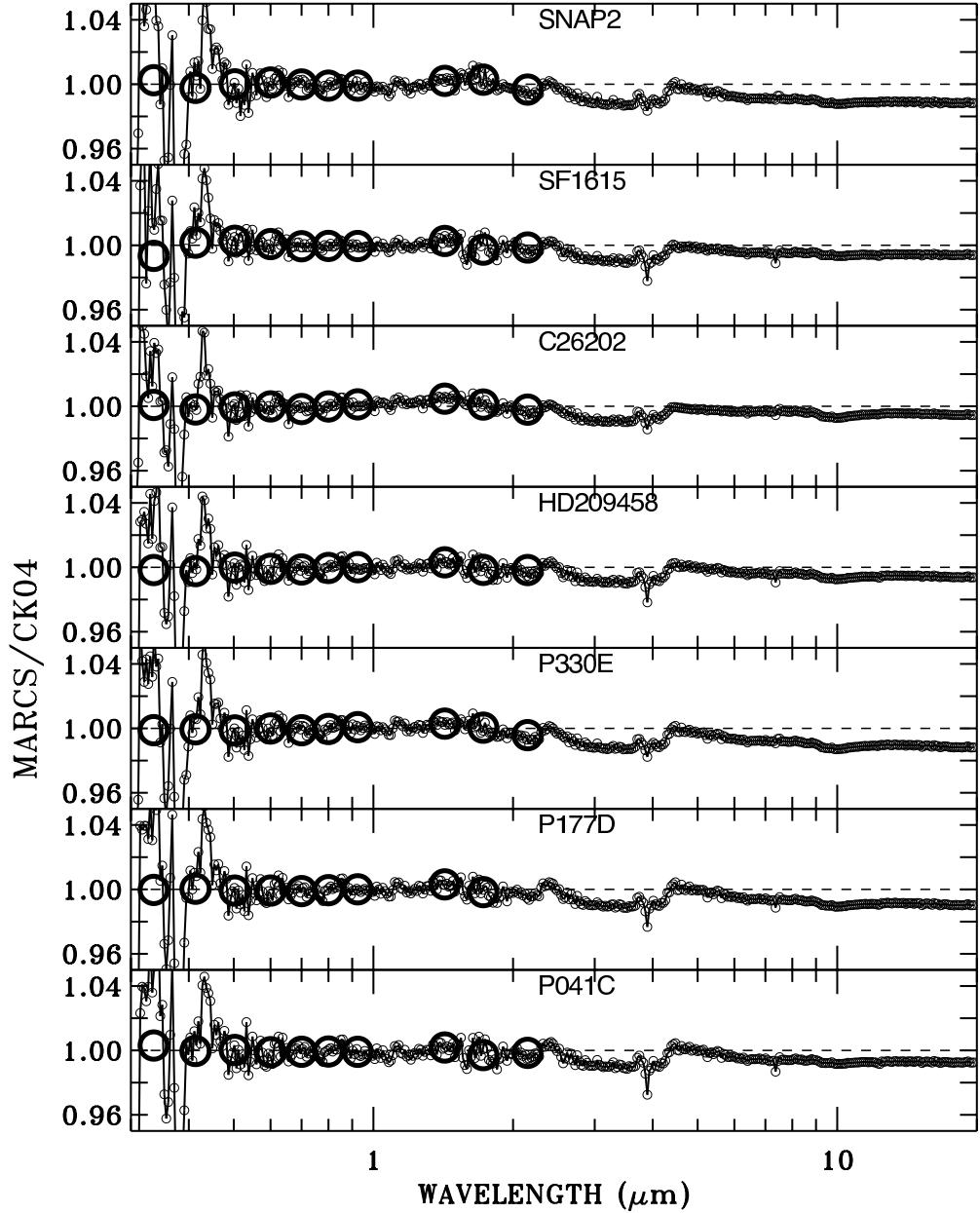}
\caption{\baselineskip=16pt
Ratios of the best-fit MARCS models to the best fitting CK04 models at $R=100$ (\emph{small open circles}) for the seven \emph{HST} standard stars. The \emph{large, thick open circles} are the ratios in the broad bands of Table~1 that are used to derive the best fits. The two sets of best-fit models agree to $\sim$1\% in the continuum, which is a measure of the uncertainty in the adopted CK04 model SEDs beyond the long wavelength 2.5~$\mu$m limit of the NICMOS measurements. The broad peaks near 4.6~$\mu$m are caused by slight differences in the strength of the CO fundamental band, which is the most prominent solar feature in the mid-IR. The MARCS models do not extend longward of 20~$\mu$m; but the flatness of the ratio from 10--20~$\mu$m attests to the validity of the CK04 grid from 20--30~$\mu$m.}
\end{figure}

\begin{figure} 
\centering 
\includegraphics[height=6.5in]{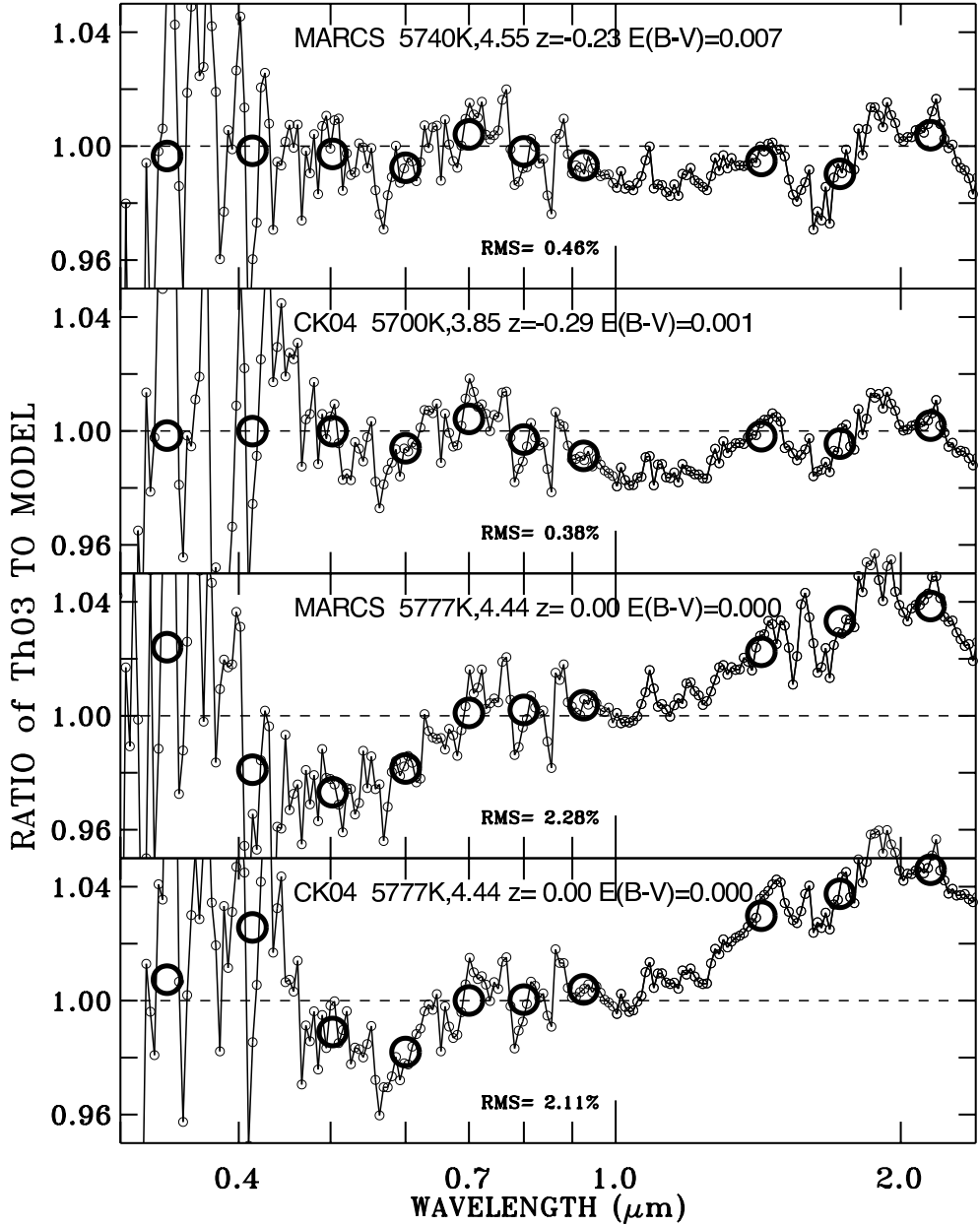}
\caption{\baselineskip=16pt
\emph{Lower two panels}: Ratio of the solar flux from Th03 to the CK04 and to the special solar model from the MARCS website. \emph{Upper two panels}: Ratio of Th03 solar SED to the best fits from the CK04 and MARCS grids, as in Figures 2--3. These ratios also correspond to comparisons with the solar flux of Rieke et~al.\ (2008), who have adopted the Th03 fluxes below 2.4~$\mu$m. The models are all normalized to Th03 over the 0.6--0.9~$\mu$m range; and the model parameters $T_\mathrm{eff}$, $\log g$, $\log z$, and $E(B-V)$ are all indicated in each panel. The models from the grids have $v_{\rm turb}=2$~km/s, while in the second panel from the bottom the special MARCS solar model has $v_{\rm turb}=1$~km/s.}
\end{figure}

\begin{figure}
\centering
\includegraphics[height=6.5in]{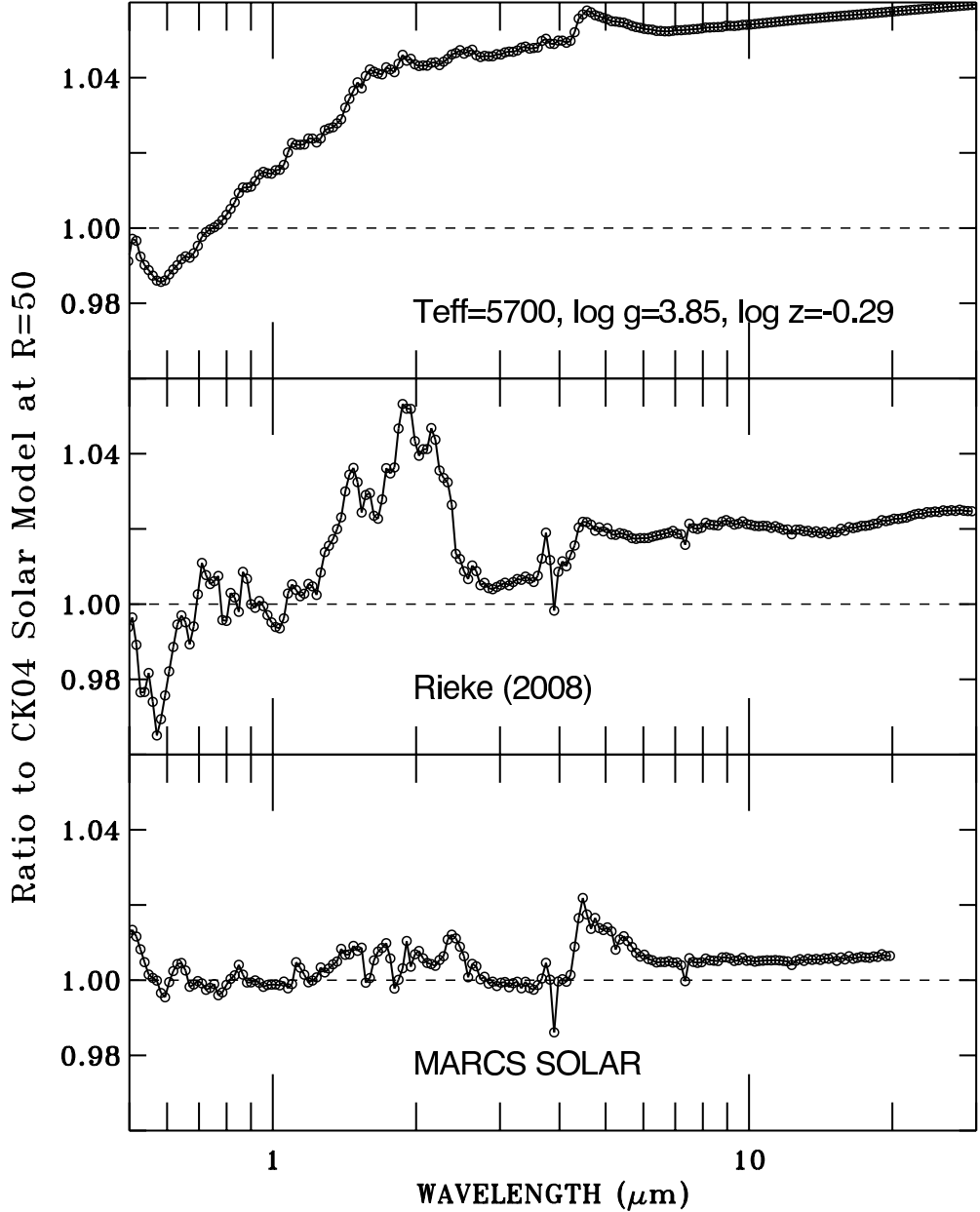}
\caption{\baselineskip=16pt
Ratio of three candidate solar SEDs to the baseline CK04 with $T_\mathrm{eff}=5777$~K, $\log g=4.44$, and $\log z=0$. The ratios are normalized to unity at 0.6--0.9~$\mu$m, have been first binned to a width of 0.02~$\mu$m, and then are re-binned to a resolution of $R=50$. The three candidate
SEDs in the numerators are: \emph{Bottom panel}:  The special MARCS solar models with $v_{\rm turb}=1$~km/s. \emph{Middle~panel}: The SED of Rieke et~al.\ (2008). \emph{Top panel}: The CK04 model SED with  $T_\mathrm{eff}=5700$~K, $\log g=3.85$, and [M/H]\,$=\log z=-0.29$ that best fits the Th03 flux distribution.}
\end{figure}

\end{document}